\documentstyle[12pt,epsf,epsfig,amsmath]{article}
\newfont{\thiplo}{msbm10 scaled\magstep 2}
\newfont{\gothic}{eufb10 scaled\magstep 2}
\newfont{\unc}{eurb10} 
\newskip\humongous \humongous=0pt plus 1000pt minus 1000pt
\def\caja{\mathsurround=0pt}\def\eqalign#1{\,\vcenter{\openup1\jot \caja
        \ialign{\strut \hfil$\displaystyle{##}$&$
        \displaystyle{{}##}$\hfil\crcr#1\crcr}}\,}
\newif\ifdtup

\def\eqright #1\cr{\noalign{\hfill$\displaystyle{{}#1}$}}
\def\eqleft #1\cr{\noalign{\noindent$\displaystyle{{}#1}$\hfill}}

\def\oldreffmt#1{\rlap{[#1]} \hbox to 2\parindent{}}

\def\figfmt#1{\rlap{Figure {#1}} \hbox to 1in{}}

\def\sectioneq{\def\theequation{\thesection.\arabic{equation}}{\let
\holdsection=\section\def\section{\setcounter{equation}{0}\holdsection}}}%

\newcounter{holdequation}

\def\begineq #1\endeq{$$ \refstepcounter{equation}\eqalign{#1}\eqno
	(\theequation) $$}
\def\contlimit{\,{\hbox{$\longrightarrow$}\kern-1.8em\lower1ex
\hbox{${\scriptstyle (a\rightarrow0)}$}}\,}
\def\centeron#1#2{{\setbox0=\hbox{#1}\setbox1=\hbox{#2}\ifdim
\wd1>\wd0\kern.5\wd1\kern-.5\wd0\fi
\copy0\kern-.5\wd0\kern-.5\wd1\copy1\ifdim\wd0>\wd1
\kern.5\wd0\kern-.5\wd1\fi}}
\def\centerover#1#2{\centeron{#1}{\setbox0=\hbox{#1}\setbox
1=\hbox{#2}\raise\ht0\hbox{\raise\dp1\hbox{\copy1}}}}
\def\centerunder#1#2{\centeron{#1}{\setbox0=\hbox{#1}\setbox
1=\hbox{#2}\lower\dp0\hbox{\lower\ht1\hbox{\copy1}}}}
\def\lsim{\;\centeron{\raise.35ex\hbox{$<$}}{\lower.65ex\hbox
{$\sim$}}\;}
\def\gsim{\;\centeron{\raise.35ex\hbox{$>$}}{\lower.65ex\hbox
{$\sim$}}\;}

\def\super#1{\ifmmode \hbox{\textsuper{#1}}\else\textsuper{#1}\fi}
\def\textsuper#1{\newcount\holdspacefactor\holdspacefactor=\spacefactor
$^{#1}$\spacefactor=\holdspacefactor}

\def\getcite#1,{\advance\citenumber by1
\def\getcitearg{#1}\def\lastarg{@}
\ifnum\citenumber=1
\ref{#1}\let\next=\getcite\else\ifx\getcitearg\lastarg\let\next=\relax
\else ,\ref{#1}\let\next=\getcite\fi\fi\next}

\def\pom{{\rm P\kern -0.53em\llap I\,}}
\def\spom{{\rm P\kern -0.36em\llap \small I\,}}
\def\sspom{{\rm P\kern -0.33em\llap \footnotesize I\,}}

\relax
\def\contlimit{\,{\hbox{$\longrightarrow$}\kern-1.8em\lower1ex
\hbox{${\scriptstyle (a\rightarrow0)}$}}\,}
\def\upon #1/#2 {{\textstyle{#1\over #2}}}
\relax
\renewcommand{\thefootnote}{\fnsymbol{footnote}} 

\def\mainhead#1{\setcounter{equation}{0}\addtocounter{section}{1}
  \vbox{\begin{center}\large\bf #1\end{center}}\nobreak\par}
\sectioneq
\def\subhead#1{\bigskip\vbox{\noindent\bf #1}\nobreak\par}

\def\sumtil{\centeron{\hbox{$\displaystyle\sum$}}{\lower
-1.5ex\hbox{$\widetilde{\phantom{xx}}$}}}

\def\l{\large}

\begin{document} 

\begin{titlepage} 

\rightline{\vbox{\halign{&#\hfil\cr
&\today\cr}}} 
\vspace{0.5in} 

\begin{center} 
  
{\large\bf The ``Crisis in Fundamental Physics''

- Will the LHC Pomeron End it~?~\footnote{Presented at the 5th 
Manchester Forward Physics Workshop, Dec.~2007.}}

\vspace{0.25in}

Alan. R. White\footnote{arw@hep.anl.gov } 

\vskip 0.6cm
 
\centerline{Argonne National Laboratory}
\centerline{9700 South Cass, Il 60439, USA.}
\vspace{0.5cm}

\end{center}

\begin{abstract} 
 
SU(5) gauge theory
with massless left-handed fermions in the representation
${\bf 5\oplus 15\oplus 40\oplus 45^*}~$ {\bf (QUD)} may have a bound-state
S-Matrix that, uniquely, contains
the asymptotically unitary Critical Pomeron and which
might also reproduce the full Standard Model.
If so, QUD would provide an underlying unification 
for the Standard Model in which very similar massless fermion anomaly 
dynamics is responsible for the physics of   
the strong and electroweak interactions and all particle masses are generated dynamically.
The color sextet quark sector, responsible for both
electroweak symmetry-breaking and dark matter in QUD, is predicted to produce
large cross-section effects at the LHC, with the pomeron as
a vital diagnostic - via TOTEM/CMS and FP420.

In this talk, the multi-regge construction 
of high-energy QUD amplitudes is outlined
as is, briefly, the LHC pomeron physics. Surrounding motivational 
issues (particularly outstanding QCD problems) and consequences
are also discussed. The S-Matrix anomaly physics
is conceptually and philosophically radical
with respect to the current theory paradigm. As a consequence, QUD 
could provide a welcome way out of
the current ``Crisis in Fundamental Physics'' with, potentially, it's
novel physical applicability resolving a variety of Standard
Model and, perhaps also, cosmology problems - it has zero vacuum energy.

\end{abstract}

\renewcommand{\thefootnote}{\arabic{footnote}} \end{titlepage}

\mainhead{ 1. INTRODUCTION.}

As I suggested\cite{mct} at this meeting last year, SU(5) gauge theory
with massless left-handed fermions in the representation
$5\oplus 15\oplus 40\oplus 45^*$, which I refer to as
{\bf QUD\footnote{Quantum Uno/Unification/Unitary/Underlying Dynamics}},
may have a bound-state
S-Matrix that, uniquely, contains the Critical Pomeron and which
might also reproduce the full Standard Model. According to my arguments, strikingly simple
massless fermion anomaly physics would then provide a very special 
underlying unification of the strong and electroweak interactions. All particle masses 
would be dynamical and, remarkably, although there is an underlying SU(5) gauge symmetry 
there would be no new S-Matrix interactions beyond those of the Standard Model. 

That the Critical Pomeron 
uniquely satisfies all the constraints of t- and s-channel multiparticle
unitarity could, perhaps, imply that QUD is 
unique as a field theory generating a unitary massive 
particle S-Matrix. (Most likely, only the S-Matrix is  
well-defined non-perturbatively.)
In total, the implications of QUD's existence are so far reaching that, initially,
``the willing suspension of disbelief''\cite{wsd} may be required
to even consider it. Almost certainly, experimental discovery 
will be required to produce the serious interest that I believe is merited.
Fortunately, if QUD does underly the Standard Model,
the color sextet quark sector producing both
electroweak symmetry-breaking and dark matter should be dramatically evident via 
large cross-section effects at the LHC - particularly in the double pomeron cross-section
that TOTEM/CMS and FP420 will measure.

As part of this talk\cite{t07}, 
I will outline the multi-regge construction that potentially relates
the QUD high-energy S-Matrix to the Standard Model. To carry through
the full program\cite{amtm} outlined is an enormous challenge  
that badly needs the extra participation that
experimental encouragement would surely bring.  
I will briefly review the relevant LHC pomeron
physics but will not discuss at all the related 
Cosmic Ray, Tevatron, and {\Large ${\scriptstyle S\bar{p}pS}$}
evidence that I have described elsewhere\cite{mct,arw05}.

A major focus of the talk will be on surrounding motivational
issues and consequences. In particular, why high-energy unitarity could be essential
in providing the key to new physics, will be clear from my discussion of 
existing problems in the formulation of QCD. The search for the unitary Critical
Pomeron then leads to the novel dynamics
which underlies the existence of the QUD high-energy S-Matrix - the central element
being the color compensation of perturbative reggeon exchanges
by anomalous wee gluons coupled to massless fermion anomalies. For the strong interaction
QCD sector the anomalies are infra-red, while in the electroweak sector 
the interactions remain effectively perturbative at low energy because the anomalies 
involve infinite fermion transverse momenta.
In general, the dynamics (the absence of a Higgs' field included) is both conceptually and
philosophically radical with respect to the current theory paradigm, which 
is (presumably) a major reason why QUD has not emerged in direct unification searches.
In fact, the current theory paradigm is actually in a {\it crisis mode} of 
serious self-doubt that 
QUD could provide a very welcome way out of. As I will describe, it is just 
because the physical applicability of QUD is at variance with conventional
wisdom that it could, potentially, resolve a wide range of existing puzzles in the Standard
Model (and cosmology - it has zero vacuum energy).

\mainhead{2. THE CRISIS}

Arguments by leading theorists (Weinberg\cite{swe}, Susskind\cite{lsu}, and 
others\cite{bsc}) 
have led to what Smolin\cite{lsm} has called a
{\bf ``Crisis in Fundamental Physics''}. The crisis is epitomized by asking the 
following question - based on the ``string-theory~landscape''.
\begin{center}
{\it `` With an infinity of universes 
proposed, and more than $10^{400}$ theories,
is experimental proof of physical laws still possible? ''}
\end{center}
A retreat to the {\it (end of science?)} 
``anthropic principle'' - physical parameters are
determined by the existence of life - is a common response.
Adding to the bewilderment (says Smolin),  
in the 35 years since the formulation of the Standard Model, 
all proposals for ``new physics'', including
 {\it GUTS, supersymmetry, technicolor, string theory and
(most recently) extra dimensions,}
have failed to make any contact with experiment - even while
introducing a wide variety of interactions, particles, and new parameters.
(The last discovery of a new interaction
was, of course, much more than 35 years ago!) To quote Schroer's opening line\cite{bsc}
\begin{center}
{\it ``There can be no doubt that after almost a century of impressive success
fundamental physics is in the midst of a deep crisis.''} 
\end{center}  
 
Theoretically, there are
far too many (ill-defined) candidate new theories, while 
experimentally there are none! Searching for new theories via 
the symmetry-based aesthetics championed by much of the
theory community has not found a focus, even though there has been no shortage
of imagination! Doubling the number of particles to achieve 
supersymmetry, when not a single partner has been seen,
and adding extra dimensions that ``curl up''
out of our sight both seem far-fetched to ``real world'' 
theorists, as well as many experimenters. (If neither is discovered at the 
LHC, this will surely be the historical perspective~!) Indeed, all of the theoretical
constructs that lead to the epitome question above are pure speculation for which
there is no experimental evidence. In this sense, the ``crisis'' is entirely 
self-induced by
theorists. Apparently, there is insufficient constraint (either theoretical
or experimental) on the development of new 
ideas and there is much concern\cite{bsc,lsm} that a major change may be needed
in the paradigm underlying the current formulation of particle theory.

\mainhead{3. THE PARADIGM ?}

In 1999 Gross\cite{djg} discussed how 
the discovery process that led to QCD
invalidated the preceding attempt to obtain a strong interaction theory by 
bootstrapping a (supposed to be unique) unitary S-Matrix. He stated
a commonly held opinion. 
\begin{center}
{\it ``We now know that there are an infinite number of 
consistent S-Matrices that satisfy all the sacred principles. 
One can take any non-abelian gauge theory, with any 
gauge group, and many sets of fermions ... The hope for uniqueness must wait for a 
higher level of unification.''}
\end{center} 
In fact, this statement is ``breathtakingly misleading'' -
to use Bill Clinton's
phrase\cite{bcl}.
Gauge theory S-Matrices can only be calculated perturbatively. 
In D=4 the perturbation expansion for every field 
(and string) theory is wildly divergent and, almostly certainly, can not be
summed. There is no non-perturbative
formulation of any theory that can derive S-Matrix 
amplitudes - let alone discuss unitarity.

Nevertheless, it became accepted 
that general principles, including (most particularly) non-perturbative unitarity, are 
irrelevant in the search for a physical theory. The conventional wisdom 
developed that the Standard Model is a well-defined 
lagrangian quantum field theory, since it was assumed that once
perturbative renormalizability had been 
proved all the desired properties of a full, non-perturbative,
field theory would also be satisfied\footnote{\openup-1\jot{Even 
though it was acknowledged that a mathematical effort of unimaginable, herculean, 
\newline magnitude would be required to prove this\cite{jw}.}} - with a unitary S-Matrix 
automatically included. From this viewpoint,
the Standard Model fits the experimental data but, otherwise, is just one 
of an infinity of renormalizable field theories that nature could have chosen. Progress
beyond the Standard Model would only be determined by new experimental
phenomena that would have to be similarly fitted by an enlarged field theory. 
Since such phenomena had yet 
to be discovered and long distance physics was well understood via QCD, 
they would necessarily have to occur at a short distance not yet explored. 
Given the viewpoint expressed by Gross, 
theorists speculating about new physics that might appear 
were not fettered by any fear that the long-distance physics demand for
unitarity of the physical S-matrix could make any significant selection
amongst candidate theories.

The general assumption was that there should be
a unifying quantum field theory that would include the Standard Model
(and which, if the quantization of gravity is to be 
included, should embed in a string theory
- the search for which became, of course, the major preoccupation 
of the theory community over the last two decades, 
culminating in the epitome question of the last 
Section). The unifying field theory would be detected via the required
additional interactions. The extra, far from trivial, assumption had to be made 
that there would be no conflict
between the intrinsic non-perturbative applicability of QCD (involving 
confinement) and the perturbative applicability of the electroweak sector. 
Although there was no explicit understanding of how it could happen, it was believed
that a transition from perturbative to non-perturbative confinement 
physics would simply
be a consequence of the evolution of couplings with the scale involved.

Gross\cite{djg} also argued that the regge region had been given too much 
attention\footnote{Experimenters 
working on diffraction have lived with this attitude for a long time.}.
He said that short-distance experiments provide the most information
and that regge behavior is merely
\begin{center}
{\it ``an interesting, unsolved, and complicated problem for QCD''}
\end{center}
If the arguments that lead me to QUD are correct, this widely held 
view could not be further from the truth. The regge region is where, a priori, the 
connection between perturbative and non-perturbative physics should be explicitly 
evident (since a mixture of small and large momenta is involved). Indeed, 
regge-region (reggeon) unitarity is deeply related to 
other fundamental problems in the formulation of QCD (that I will discuss below)
and is central, I will argue, in the construction of a fully unitary gauge theory.
The viewpoint that the more difficult dynamical problems 
in QCD can be put aside because they are not fundamental for going beyond the 
Standard Model, has clearly been a major factor in allowing the unlimited
speculation that has produced the theory ``crisis''. It should be noted that the 
freedom of invention associated with 
the guiding principle/paradigm that progress will come via inspired guesses for missing  
short-distance physics, combined with experimental 
verification via related rare processes, has not yet received any 
experimental confirmation and, most importantly, there is no historical precedent
for assuming that it will.

Indeed, as we discuss next, the long distance and regge region physics of QCD 
is not well understood
and our argument will be that it is by getting this physics right that we are 
led directly to the underlying physics of the full Standard Model.

\mainhead{4. QCD PROBLEMS}

There are three interconnected, unresolved, 
problems for the standard formulation of QCD.

\subhead{1. The Spectrum of States}  

The conventional wisdom is that
the physical states are determined by the two principles of
color confinement and (when quarks are involved)
chiral symmetry breaking. Neither principle has been proved, although
there is also no experimental violation of either principle.
However, if color confinement is the only feature
constraining the field theory degrees of freedom appearing in  
physical states, then glueballs should be everywhere. As Chanowitz says\cite{mch},
\begin{center}
{\it ``Glueballs are a dramatic consequence of the local, unbroken, non-Abelian
symmetry that is the unique defining property of QCD. ... 
The prediction that glueballs exist is simple and fundamental but has proven 
difficult to verify''.}
\end{center}
In fact,
\begin{center}
{\it not a single glueball 
has been seen {\bf (unambiguously)} in any experiment\cite{kz}.}
\end{center}
Apparently, there is a major limitation
on the degrees of freedom - physical states must contain quarks ?? This problem
has received only limited theoretical attention, in part because there is no  
resolution within the standard formulation of QCD. 

\subhead{2. The Parton Model} 

Factorization theorems say that leading-twist perturbation theory
is consistent with the parton model - leading to 
the ``QCD-improved parton model''. However, even though
it is the basis of all perturbative applications, there is 
\begin{center}
{\it no derivation of the parton model in QCD.}
\end{center}
As elaborated on at length in \cite{arw84}, the true parton model envisaged
by Feynman requires that infinite momentum hadrons have 
quark/gluon wave functions. This is a very intricate requirement 
that has no reason to be true 
if there is a non-trivial, confining, vacuum. (Even though it is probably essential for 
asymptotic freedom to be maximally applicable.) 

For the desired
wave-functions to exist, the ``wee partons'' - with finite momentum in the
infinite-momentum frame (and zero momentum at finite momentum) 
- must play the role of the vacuum and so must be universal.
As illustrated, schematically, in Fig.~1, the wee partons in a hadron dominate
pomeron exchange. For wee partons to be universal, the pomeron
must have the factorization properties of a regge pole.

This problem has also not received much attention, presumably because the parton model, 
like confinement, works so well that (since QCD is obviously correct!!)
it can only be that it is indeed valid in QCD.
\begin{center}
\epsfxsize=3.5in
\epsffile{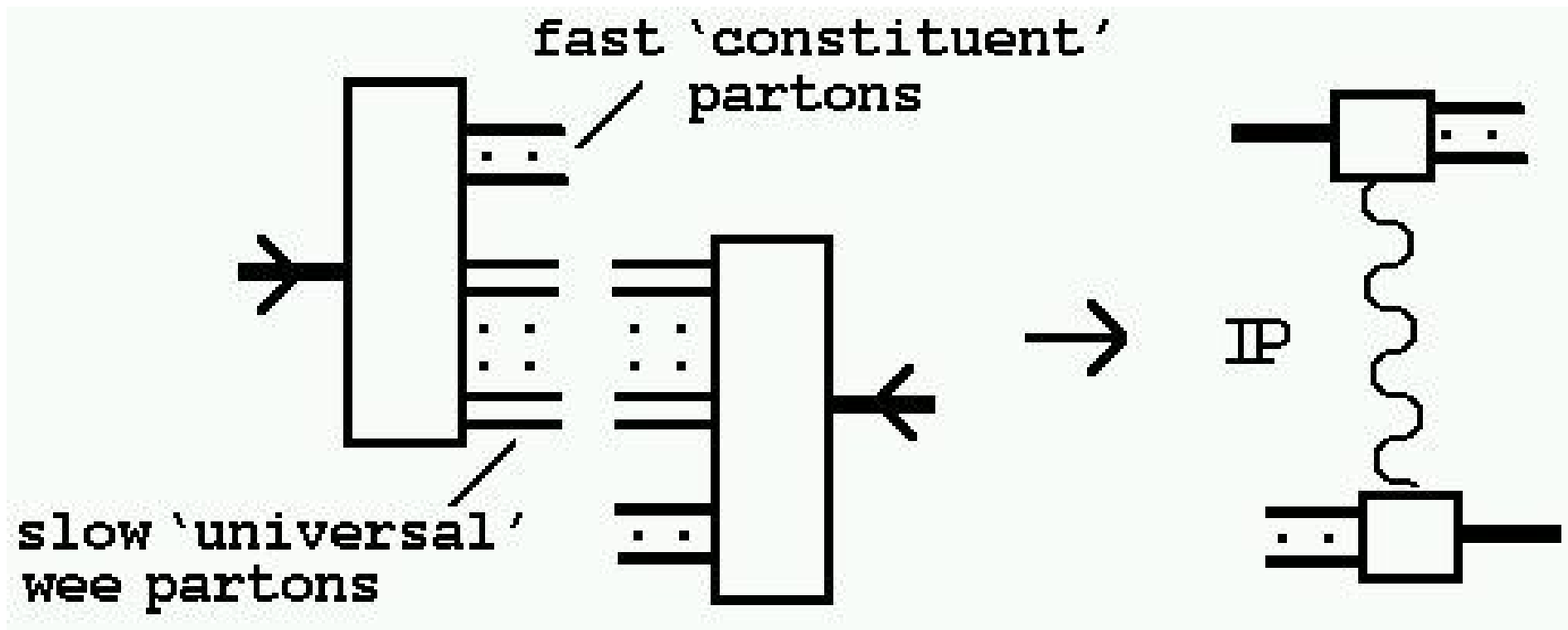}

Figure 1. Wee Partons and the Pomeron
\end{center}

\subhead{3. The Pomeron} 

A priori, the perturbative  BFKL pomeron should appear at 
large $k_{\perp}$ - together with an odderon. However,
as Ewerz says\cite{ce},
\begin{center}
{\it 
``QCD predicts the existence of the perturbative pomeron and of the odderon.
\newline Both of them appear to be rather difficult to observe experimentally.''}
\end{center}
In fact,
\begin{center}
{\it there is no {\bf (unambiguous)} 
evidence for the BFKL pomeron 
\newline and, essentially, zero evidence for the odderon.}
\end{center}
At small $k_{\perp}$, it is established experimentally that the pomeron
couples directly to quarks (c.f.~total cross-sections) 
and that it has the factorization properties of a regge pole.

Apparently, BFKL gluons (which would be directly related to a sub-set
of glueballs) could well be 
absent in physical high-energy amplitudes. Moreover,
quarks should be the essential elements of the high-energy states. The BFKL pomeron
has received so much attention and has produced such an enormous literature that there
is major resistance, by theorists, to acknowledging the
lack of experimental confirmation that is, nevertheless, well-known to experimenters. 

The BFKL pomeron also has serious theoretical problems. Firstly,
it violates both s- and t-channel multiparticle 
unitarity\cite{arw90}
(with higher-order corrections, very likely, making the problem worse).
Secondly, because it is not a regge pole,
wee parton factorization is absent and Reggeon Field Theory (RFT) can 
not be used to obtain higher-order contributions\cite{arw90,arw91} via reggeon unitarity.

That the gluonic degrees of freedom involved in the physical states and amplitudes of
QCD are more restricted than conventionally assumed, and that this 
may be related to the wee parton physics underlying the parton model, is the
vital, conceptually radical, message that will take us beyond the Standard Model to QUD.
 
\mainhead{5. THE CRITICAL POMERON AND QUD}

The Critical Pomeron\cite{cri} is obtained as a fixed-point
renormalization group solution of RFT and uniquely
satisfies both $s$- and $t$-channel multiparticle unitarity. It was originally formulated
as an elementary regge pole plus interactions for two reasons. Firstly, as noted above,
the pomeron is approximately a regge pole phenomenologically
and, secondly, $t$-channel reggeon unitarity is satisfied\cite{arw91} by the build-up of
the RFT interactions of a regge pole. As we noted in the last Section, a third reason 
is provided by requiring the factorization 
property needed for the existence of a parton model 
- for which the pomeron must be a {\bf single} regge
pole (plus interactions).  

In a gauge theory, and QCD in particular, reggeon diagrams (transverse momentum
diagrams containing ``reggeized'' gauge boson and fermion regge pole propagators) 
provide a well-established perturbative description of high-energy regge behavior.
If a QCD parton model can be derived then, a priori, 
these diagrams might be expected to provide the appropriate technology.
An immediate problem is that if confinement is to appear, it  
can only be via infra-red divergences. That
the Critical Pomeron, along with the wee partons required for 
the parton model, can indeed be obtained
from infra-red divergent reggeon diagrams is a highly non-trivial
construction that it has taken me many years to formulate. 
As I describe below, a very special version of QCD (QCD$_S$) 
is required, which then has to be embedded in a unique 
larger theory - QUD. Confinement is indeed part of the outcome, but there is a 
much stronger restriction, than simple zero color, on the gluon degrees of freedom.

It might be anticipated that a major restriction 
on field theory degrees of freedom would involve fermion anomalies. Moreover, 
only the anomalies of massless fermions have infra-red significance.
As we will see, it is novel massless fermion anomaly dynamics that strongly
restricts the physical degrees of freedom in both QCD$_S$ and QUD. Longitudinal 
gluon interactions allowed\cite{arw84} by the Gribov quantization ambiguity also play
a crucial role. My uncovering of this dynamics 
relates back to the controversy\cite{scrft} that occurred thirty years ago 
concerning the properties of a unitary supercritical RFT phase.
To satisfy reggeon unitarity
in this phase, a ``pomeron condensate'' has to be produced by the wee partons
in a hadron\cite{arw91}. Obtaining
this condensate from the infra-red divergences
of reggeon diagrams, is the key
to understanding the origin of the Critical Pomeron in both QCD$_S$ and QUD.
 
Before describing the dynamics, we first summarize the path that connects
the Critical Pomeron to QUD and describe the basic properties of this theory.

\newpage 
\subhead{5.1 The Path to QUD}

{\it \begin{enumerate}
\item{The first step is to match\cite{arw05} supercritical RFT with the 
superconducting phase of SU(3) gauge theory  
in which the color symmetry is reduced to SU(2).} 
\item{The next step is to show that the Critical Pomeron 
occurs in unbroken QCD when asymptotic freedom is ``saturated''. 
This is achieved with six color triplet quarks
plus two color sextet quarks - giving what we call ``QCD$_S$''.}
\item{Adding the electroweak sector, 
the $W^{\pm}$ and $Z^0$ eat the ``sextet pions'' 
and electroweak symmetry breaking occurs\cite{wm}
- with no new interaction! The electroweak scale is the QCD
sextet chiral scale - in agreement with Casimir scaling!}
\item{To cancel the electroweak short-distance anomaly and to generate masses, the sextet 
sector is embedded in a left-handed unified theory. Asking\cite{kw}
for the appropriate electroweak quantum numbers for the sextet sector, 
together with asymptotic freedom and anomaly cancelation, {\l \bf uniquely} selects
SU(5) gauge theory with the fermion representation
\begin{center} ${\l \bf 5\oplus 15\oplus 40\oplus 45^* ~\leftrightarrow ~QUD} $
\end{center}}
\end{enumerate}}

\subhead{5.2 SU(3)xSU(2)xU(1) properties of QUD}

QUD contains QCD$_S$ and under 
$~SU(3)\otimes SU(2)\otimes U(1)$ the SU(5) representations give
$$
{\openup-1\jot 
\eqalign{5&=(3,1,-\frac{1}{3})^{ \{3\}}
+(1,2,\frac{1}{2})^{\{2\}}~,\cr
 15&=(1,3,1)+
(3,2,\frac{1}{6})^{ \{1\}}+{\bf (6,1,-\frac{2}{3})}~,\cr
40&=(1,2,-\frac{3}{2})^{ \{3\}}
+(3,2,\frac{1}{6})^{ \{2\}}+ (3^*,1,-\frac{2}{3})~+~(3^*,3,-\frac{2}{3})\cr
&~~~+
{\bf (6^*,2,\frac{1}{6})}+(8,1,1)~,\cr
45^*&=(1,2,-\frac{1}{2})^{\{1\}}+(3^*,1,\frac{1}{3})
+(3^*,3,\frac{1}{3})+(3,1,-\frac{4}{3})
+(3,2,\frac{7}{6})^{\{3\}}\cr
&~~~+
{\bf (6,1,\frac{1}{3})}+(8,2,-\frac{1}{2})}}
$$
The requested sextet sector is shown in bold face.
The triplet quark and lepton sectors, {\it which were not asked for,} are 
amazingly close to the Standard Model ! There are three ``generations'', 
labeled by \{1\},\{2\} and \{3\}. 
The electroweak $SU(2)\otimes U(1)$ quantum numbers are not quite right
and so to be physically relevant this sector, like the QCD sector,
can not be perturbative.
After the initial discovery\cite{kw} of QUD it was very frustrating
that no form of symmetry breaking appeared able to give the Standard Model directly.
Only after I fully understood the anomaly dynamics producing the QCD$_S$ S-Matrix, 
did I realise
that QUD could give the Standard Model S-Matrix via the same dynamics. 
For this, it is very important that the lepton anomaly is correct
and that there are three sets of triplet quark/antiquark pairs with charges 
$\pm \frac {2}{3}$ and $\pm \frac{1}{3}$. It is also crucial that, as can be seen
from the above representation decompositions, QUD is charge conjugate, and therefore
vector-like, only with respect to $SU(3)\otimes U(1)_{em}$.

It is surely remarkable that by simply asking for a sextet quark sector capable of
producing electroweak symmetry breaking, we are led to a minimal SU(5) theory that, via the
dynamics that I outline next, contains
just enough, and only enough, to produce all the states and interactions of the Standard
Model. As we will see, beyond the potential quark and lepton generations there is only
\begin{enumerate}{\it
\item{A sextet quark sector that, as a complex SU(3) representation
with a large color casimir, produces a new large mass strong interaction sector that is 
responsible for both electroweak symmetry breaking and dark matter.}
\item{A ``lepton-like'' octet quark sector that, as a real SU(3)
representation, provides a short-distance anomaly contribution which produces
SU(5) invariant leptons that have no strong interaction and that, together with similarly 
produced hadrons, form Standard Model generations.}
\item{Two exotically charged quarks that, apparently, have no dynamical
significance.}}
\end{enumerate}

\mainhead{6. STATES AND AMPLITUDES}

A priori, obtaining not only bound states but also their scattering amplitudes,
would be considered an impossible task in a gauge theory where there are (according
to our own assertion) no off-shell Green's functions. Fortunately, 
as we outline below, multi-regge theory (very specially)
allows us, in principle at least, to simultaneously construct bound-states and S-Matrix 
amplitudes, for both QCD$_S$ and QUD, using perturbative reggeon diagrams. 
Since multi-regge limits are defined, in effect, so that 
infinite momentum frame kinematics are  
introduced in some Lorentz frame for each state and interaction, in the special case
that ``vacuum properties'' are carried by ``universal wee partons'' as discussed
in Section 4, a perturbative starting point may be sufficient to
access multi-regge region ``non-perturbative'' physics.
In fact, this physics will emerge in a surprisingly simple manner. 

Although the ``infinite momentum'' description of physical states is likely to 
be quite different from descriptions at finite momentum there must, of course, 
be consistency.
In fact, since the infinite momentum (or regge) 
limit is where perturbative and non-perturbative physics must come together, the form
of the infinite momentum states can be viewed as a boundary condition determining
whether or not candidate bound-states, that appear to be present at finite momentum,
can be present when there is a matching with large momentum perturbation theory. 

The multi-regge formalism 
also allows us to systematically build up 
the effects of massless fermion anomalies at high energy. (We will not attempt to 
discuss what the corresponding finite energy phenomena might be.) As we will describe,
we start in a color-superconducting phase  in which all 
reggeons are massive (i.e the gauge symmetry is spontaneously broken - completely). 
The reggeons are also gauge-invariant
and carry global representations of the gauge group. It is well-known that
in the massless limit, the reggeization exponentiation of infra-red 
divergences ``confines'' the global color - although color zero massless reggeon states
composed of massless gluons remain. In our construction, these states are removed
and true confinement is produced by an additional, 
non-exponentiating, ``anomaly divergence''. This divergence will also be the source
of the pomeron condensate that we are looking for.

There are two elements that combine to produce the anomaly divergence. 
The first is ``anomalous gluons'', which are configurations of 
gauge boson reggeons that (in a limited sense) are 
multi-reggeon generalizations of the well-known ``anomaly current''.
The second is ``reggeon vertex anomalies'' which occur in fermion loop effective
vertices and provide the couplings for anomalous gluons. As we will describe, 
when there is a transverse momentum cut-off an infra-red anomaly 
divergence occurs when the transverse momenta of all anomalous gluons in a 
multi-regge amplitude are scaled uniformly to zero. In the residue of the divergence,
therefore, all reggeons are actually on mass-shell
and carry zero transverse momentum. Consequently, they  
can be referred to simply as ``anomalous wee gluons''.

\subhead{6.1 Anomalous Gluons}

For vector reggeons, the signature ($\tau$) of a reggeon state
is simply the odd/even number of reggeons.
Signature is also related to charge conjugation (C) and parity (P).
When parity is conserved, the combination of incoming and outgoing
states to which a reggeon combination couples can be assigned a parity which is also
carried by the reggeon state. In this case $\tau$ is also the sign obtained by
a TCP transformation of the complete coupling. If the scattering particles are scalars,
then T simply 
interchanges the ingoing and outgoing particles, and so it must be that $\tau=CP~$. 
Initially, we define C and P perturbatively so that C includes color charge
conjugation.

``Anomalous gluons'' are sets of gluon reggeons that carry color zero 
but have color charge conjugation 
parity C with $C\neq \tau$, implying that (when parity is conserved)
anomalous gluon couplings must carry $P=-1$. The absence of a d-tensor implies 
that for SU(2) color only odd signature anomalous gluons are possible. Also, for forward 
scattering $P= -1$ implies that there must be a 
perturbative\footnote{The ``vacuum'' presence of 
anomalous wee gluons in physical states, that we will describe, will modify 
the definition of charge conjugation and parity in the physical
S-Matrix.} parity change between the initial and the final scattering state. 
In a helicity-conserving massless 
vector theory, such a change can only come from
an anomaly vertex that contains a zero momentum chirality transition
-as we discuss next.

\subhead{6.2 Reggeon Vertex Anomalies}

Anomalies appear in ``effective triangle diagram''  
reggeon vertices that are generated when fermions in large 
loops are placed on-shell by a multi-regge limit. Examples, derived in my
papers, are shown in Fig.~2. 
\begin{center}

\epsfxsize=5.2in
\epsffile{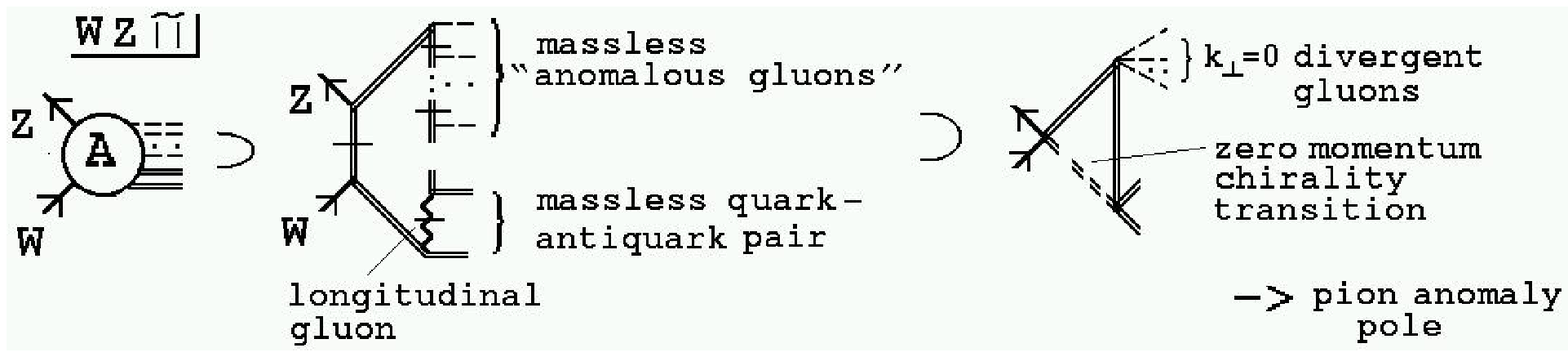}
\vspace{0.025in}
\epsfxsize=5in
\epsffile{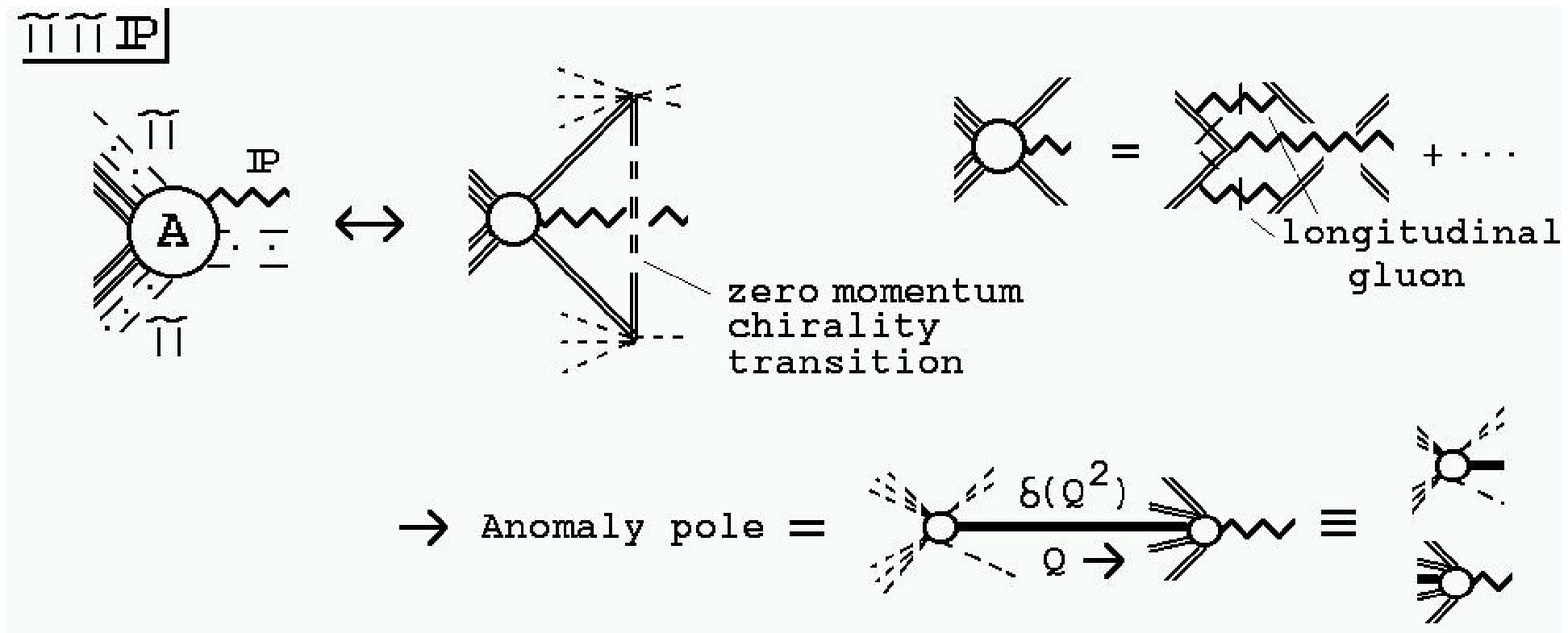}

\vspace{0.025in}

 
\epsfxsize=4.3in
\epsffile{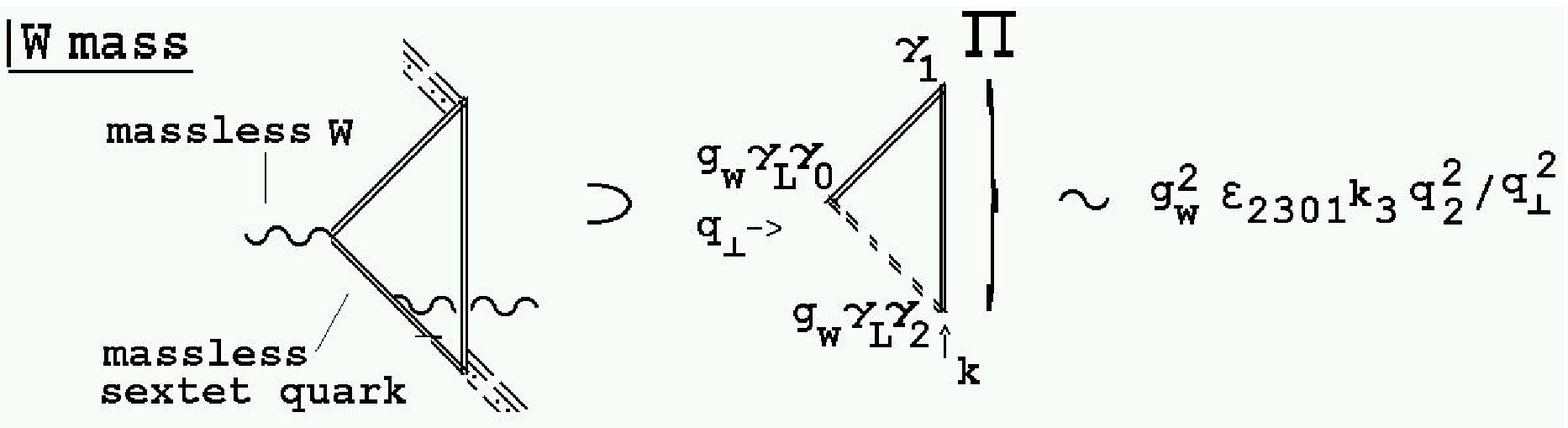}
$~~~~~~$

Figure.~2 Reggeon Vertex Anomalies
\end{center}
The longitudinal gluon interactions shown are 
present in the color superconducting phase and their survival in the massless theory 
is allowed by the Gribov ambiguity. The anomalies occur only in
vertices that couple distinct reggeon channels. As a result,
when anomalous gluons are a sub-component of a multi-reggeon state they can
not have a zero transverse momentum perturbative reggeon interaction 
with the other (massive) reggeons in the state. This property prevents
the exponentiation of the anomaly divergence in a vector theory.

A priori, the effect 
of the anomalies is dependent on the color symmetry restoration and cut-off
removal procedure used. (The procedure we employ has the
major justification that it  
produces a unitary RFT description of the high-energy behavior
by exploiting the wee parton Gauss law ambiguity.)
It is the Ward identity violation produced by the $k_{\perp}$ cut-off in anomaly
vertices that causes the infra-red anomalous gluon divergence. 
It is crucial that the residue
of the divergence contains 
anomaly poles that are generated\cite{arw05} (within the anomaly vertices) 
by zero momentum chirality transitions. 

Anomaly poles have two distinct roles.
Poles associated with a flavor anomaly are Goldstone boson particle 
poles\footnote{The kinematics responsible for
the appearance of a Goldstone boson as an anomaly pole
at infinite momentum are detailed in \cite{arw02}.}.
As illustrated in the first example of Fig.~2, 
a Goldstone (pion) pole is produced when a set of anomalous ``wee'' gluons, with
$k_{\perp}=0$, couples via an effective triangle diagram to 
a quark-antiquark pair that also carries a light-like momentum. 
Poles associated with the U(1) anomaly do not contribute as particle poles but instead
contribute as $\delta$-functions that conserve the wee gluon transverse momenta
involved in anomaly divergences in distinct reggeon channels - as illustrated 
in the second example of Fig.~2. 

A chirality transition is essential for the generation of an anomaly pole. In effect,
these transitions, and the quantum numbers of the amplitudes in which they
appear, are a consequence of 
the initial reggeon mass generation. Since they are also, essentially, the
zero momentum propagator contribution to a condensate, they 
play an analagous role to condensates - {\it but only in the
S-Matrix}. In QCD$_S$ the chirality transitions do not conflict with
the global color symmetry. Instead, they break both color parity - producing 
a regge pole pomeron and confinement - and chiral symmetry. 
In QUD only the vector part of the color 
group is not in conflict with the chirality transitions and this is why Standard Model
interactions emerge.

A comprehensive and detailed analysis 
of the momentum and color structure of reggeon vertex anomalies is 
amongst the many areas where much more work will be needed even 
after we elaborate, in \cite{amtm}, the technical details behind the outline 
we give in the following. 
It is important that the chirality transitions appear only in the 
anomaly interactions and so they do not produce the range of off-shell phenomena that
condensates produce. In fact, because the reggeon masses decouple 
straightforwardly in all non-anomaly reggeon interactions,
the associated large $k_{\perp}$ perturbation theory is given by the massless theory.
Provided, therefore, that we {\it limit our discussion to the bound-state S-Matrix}
we can combine infra-red (effective) symmetry breaking via chirality
transitions with asymptotically-free, massless, perturbation theory. 

\subhead{6.3 Multi-Regge Amplitudes Via the Anomaly Divergence}

For both QCD$_S$ and QUD we start, as we already noted, with masses for 
all reggeons (both gauge bosons and fermions). With a $k_{\perp}$ cut-off in place, we can 
smoothly restore the 
gauge symmetry in steps, utilising complimentarity\cite{fs}, provided the masses
are generated by expectation values for fundamental representation scalars.
For QCD$_S$, SU(3) color is restored via the two steps
\begin{center}
$\to SU(2)~ \to SU(3)$
\end{center}
while to reach the 
SU(5) color symmetry of QUD requires the sequence of limits
\begin{center}
$\to SU(2)~ \to SU(3) ~\to SU(4)~ \to SU(5)$
\end{center}
(In fact, we go straight from SU(2) to SU(4).)
In both cases, we remove the $k_{\perp}$ cut-off before the last step 
(utilising the scalar field asymptotic freedom that results from
a very small $\beta$-function).

As the initial SU(2) symmetry is restored, bound-states and interactions emerge  
together in multi-regge amplitudes containing the anomaly divergence. The
simplest examples are ``di-triple-regge'' amplitudes\cite{arw91}
of the form illustrated in Fig.~3.
\begin{center}
\epsfxsize=3.4in
\epsffile{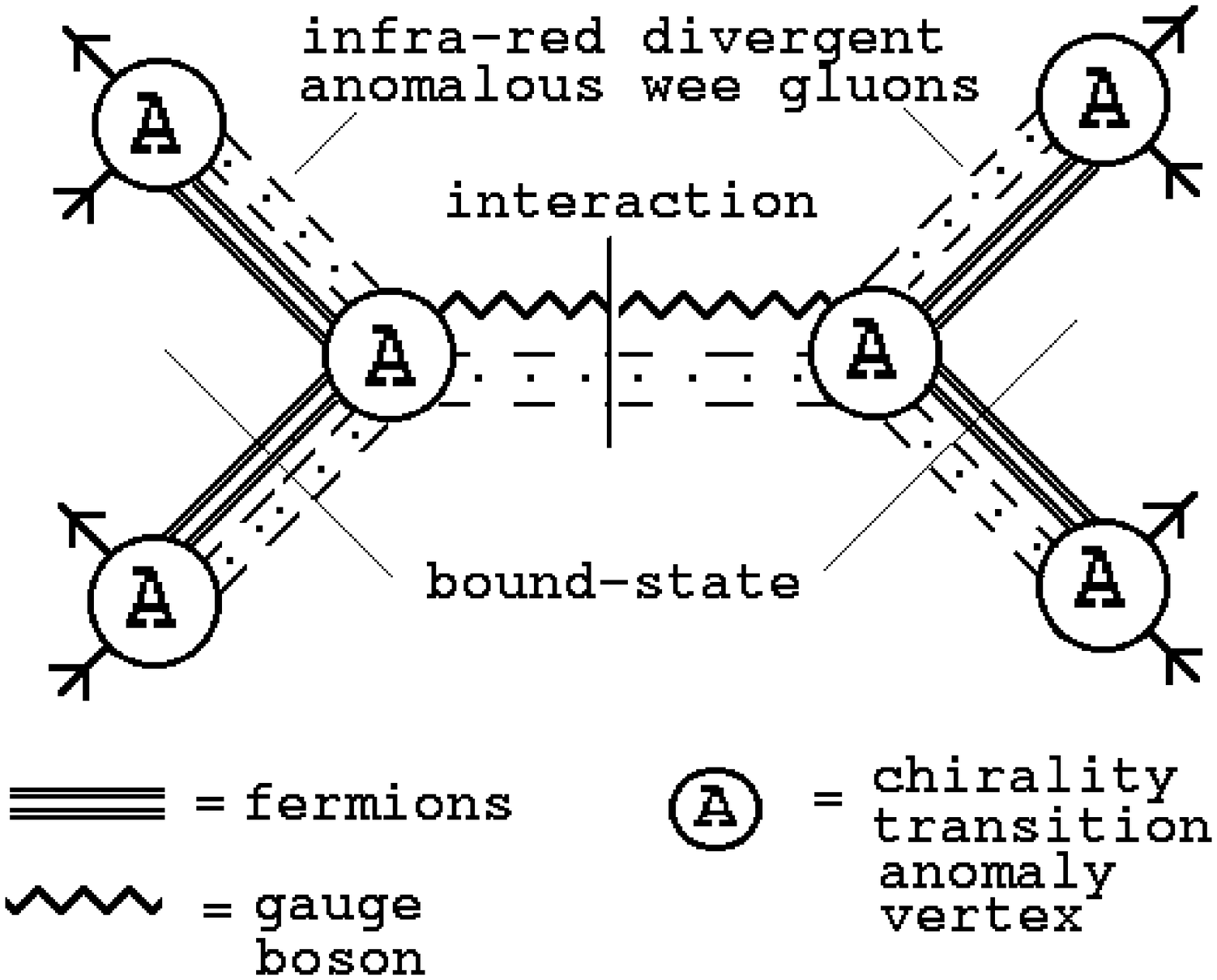}

Figure.~3 A Di-Triple-Regge Amplitudes.
\end{center} 
Because the gluons involved are anomalous, 
there are no interactions to exponentiate the
divergence. The divergence will also survive\cite{arw02} self-interactions
amongst the anomalous gluons if color zero reggeon interactions have the scaling property
that holds when all fermions are massless and there is an infra-red fixed-point\cite{bz}, 
as is vitally the case in both massless QCD$_S$ and QUD.

Physical amplitudes are obtained by 
factoring off the anomaly divergence and interpreting it as 
an odd-signature anomalous wee gluon ``vacuum condensate'' that is universally
present in all states and interactions.
The amplitude given by Fig.~3 can be represented
diagramatically as in Fig.~4. 
\begin{center}
\epsfxsize=6in
\epsffile{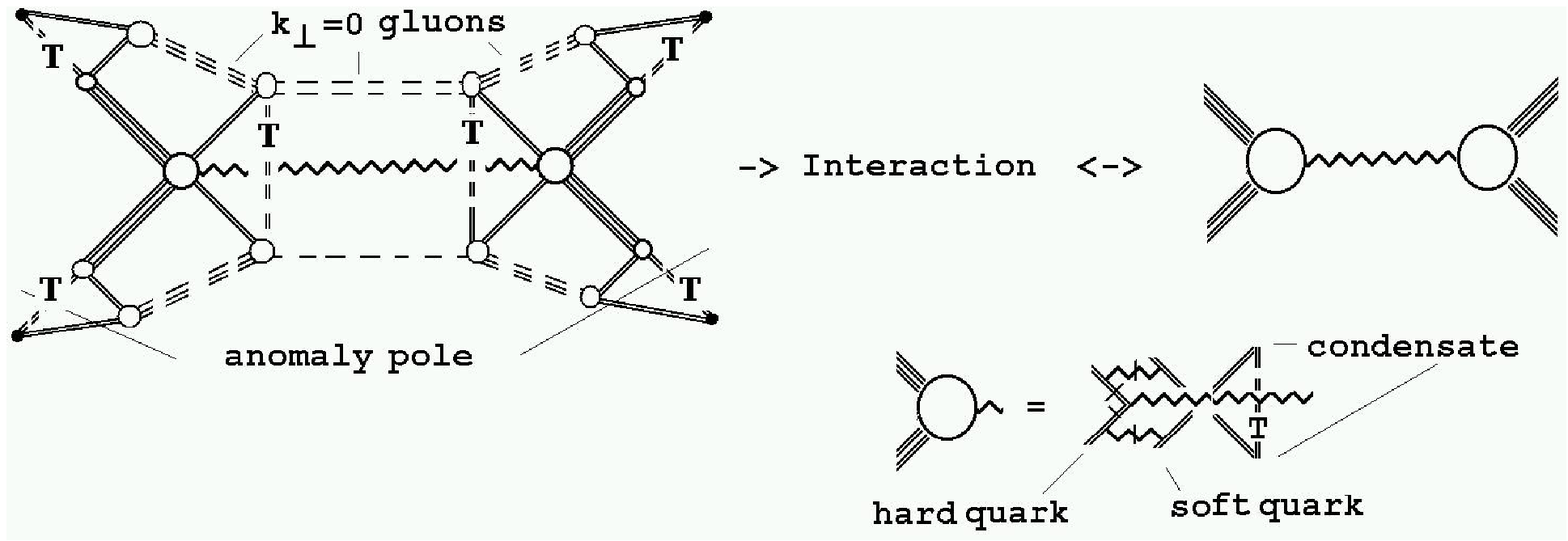}

$~$ \newline
Figure.~4 The Physical Amplitude, $T=$ Chirality Transition. 
\end{center}
The interaction that appears 
is simply the color zero combination of a finite transverse 
momentum gauge boson reggeon in the $k_{\perp}=0$ wee gluon condensate, but with  
an anomaly interaction coupling that contains a chirality transition as shown. This is
the origin of the pomeron in QCD$_S$ and of the pomeron and the 
electroweak interactions in QUD. The effect of the anomaly coupling is to produce
an even signature pomeron from an odd signature perturbative reggeon. 
At this stage, electroweak interactions are also even signature. 
They will become odd-signature as the full gauge symmetry is restored in QUD.

As is also illustrated in Fig.~4, the bound-states are 
anomaly poles that first appear after the initial SU(2) color restoration.
An anomaly pole will survive interactions if it is a Goldstone   
boson associated with the breaking of a chiral symmetry involving complex conjugate
representations of the color group. The bound-states can also be represented  
as color zero fermion/antifermion combinations in the anomalous wee gluon condensate. 
Within the effective triangle diagram the anomaly pole results from 
the production of a fermion pair, one of which is a zero
momentum hole state that undergoes the chirality transition to become physical
via the compensating anomalous wee gluon emission. 
Even though it becomes physical, the hole state remains ``soft'' while the initially 
physical fermion carries\cite{arw02} all the longitudinal momentum.  
As illustrated in Fig.~4, the hard
fermion then couples to the perturbative reggeon that carries the interaction
transverse momentum. 

According to this last description, the full bound-state interaction can be viewed as a 
hard perturbative interaction that is accompanied by a color-compensating 
wee parton ``vacuum condensate'' component whose coupling is responsible for
a ``zero-momentum shift of the Dirac sea'' in the scattering bound-states. 
Equivalently, we could drop the reference to the vacuum condensate 
exchange and instead talk only about a perturbative 
reggeon exchange that has ``vacuum produced''
color-compensating, shifts of the Dirac sea within the couplings 
so that the ``perturbative'' interaction involves zero color exchange.

\subhead{6.4 The Pomeron Condensate and the Critical Pomeron}

A very important consequence of  
the presence of the wee gluon condensate in bound-states is that, as illustrated 
in Fig.~5, it is directly responsible
for the ``vacuum production'' of pomerons in higher-order amplitudes
that is equivalent to the existence of a ``pomeron condensate''. 
This is the wee parton production of
the pomeron condensate that we anticipated in Section 5 would be the clue to the
origin of the supercritical pomeron. 
Indeed, the generated amplitudes are, apparently, just
those of supercritical RFT. If the mapping onto 
supercritical RFT is complete (much remains to be done to establish that it is),
the pomeron becomes 
critical as SU(3) color is restored. The pomeron condensate vertices will be
absorbed into regular RFT vertices as the reverse of the
process described in \cite{arw91}. 
The Critical Pomeron is then even signature and 
composed entirely of anomalous gluon configurations - including
those in the condensate - that do not include BFKL gluons.
\begin{center}
\epsfxsize=4.5in
\epsffile{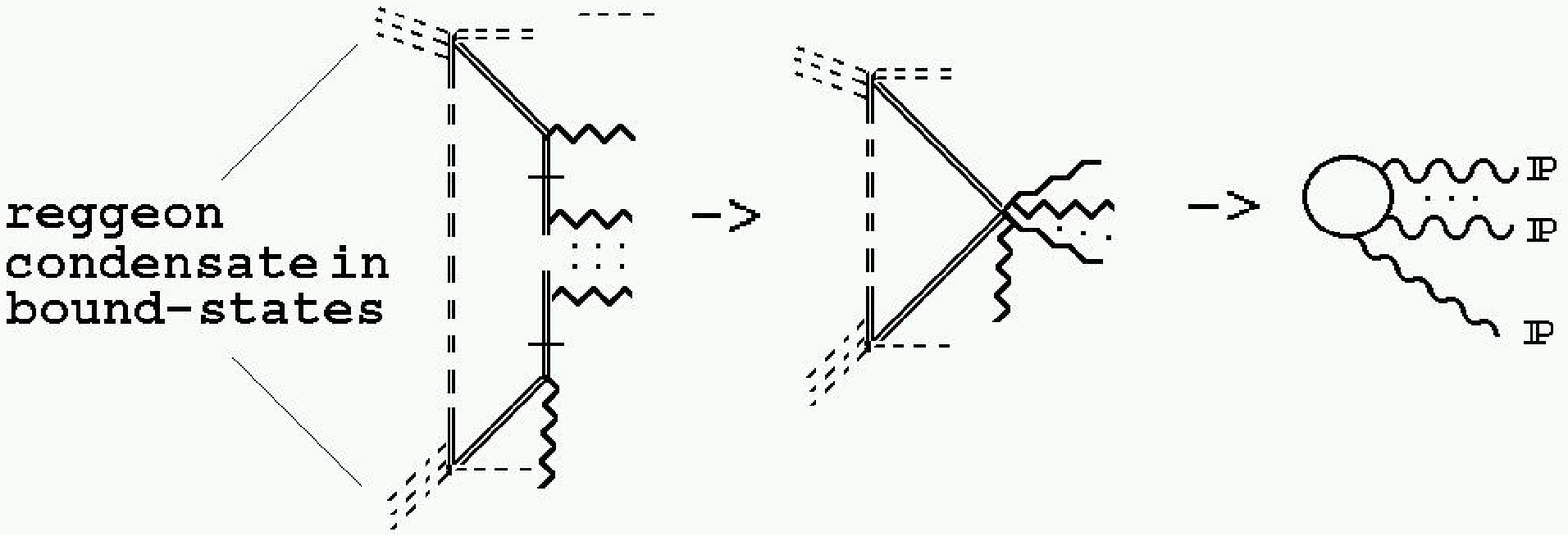}

Figure.~5 Pomeron Vacuum Vertices Produced by the Reggeon Condensate
\end{center}

Relating the anomalous gluon emission from anomaly vertices to the (hole)
displacement of the Dirac sea that produces the chirality transition, we can describe 
it as gluon production via a ``Fermi surface fluctuation''.
The fluctuation can then
be viewed as the order parameter of the pomeron phase-transition. 
In the supercritical phase it is a correlated wee parton
``vacuum condensate'' introduced by the color symmetry breaking. 
At the critical point the Fermi surface fluctuations 
become dynamical and uncorrelated (i.e. random within the SU(3) color group). 
These fluctuations then combine
with perturbative reggeon exchange to produce the color zero 
long-range collective phenomenon that is the Critical Pomeron. In QUD, as we discuss
further below,
the Fermi surface anomalous wee gluon emission (and Critical Pomeron behaviour) 
is limited to a non-abelian subgroup with vector-like couplings to the sea. Because QUD
is vector-like with respect to an SU(3)xU(1) subgroup only, 
the parity conserving SU(3) color strong interaction emerges.

Note that, because only global color symmetries (and not gauge symmetries) are involved
in our reggeon diagram constructions,
it is straightforward for color zero reggeon states to be formed from reggeons carrying
very distinct transverse momentum (including zero for condensate reggeons).

\mainhead{7. Massless QCD$_S$}

The physical states of QCD$_S$ correspond to the Goldstone bosons of the 
superconducting theory in which only SU(2) color is restored.
This includes both quark/antiquark mesons and\cite{kog} quark/quark ``nucleons''.
The nucleons aquire an extra quark as SU(3) color is restored. 
We have yet to provide a  description of how this takes place in reggeon diagrams, but we 
anticipate that the additional quark will be ``soft'' and similar, if not identical,
to the hole produced quark that is already present. 
This would produce the additive quark model for
total cross-sections and would also imply that both meson and nucleon reggeization
are a consequence of quark reggeization.

The only new states that appear, in addition to triplet mesons and nucleons, 
are sextet ``pions'' and ``nucleons'' ($P_6$ and $N_6$).
There are no hybrid sextet/triplet quark states and 
\begin{center}
{\l \it no glueballs}.
\end{center}
(This spectrum is consistent with, but much less than would be obtained by simply
imposing confinement 
and chiral symmetry breaking.)
If the sextet pions are eaten by $W^{\pm}$ and $Z^0$ vector bosons, the only
remaining new states are sextet nucleons. 
The $N_6$ will be stable\cite{arw05} and dominate 
ultra high-energy cross-sections. Sextet neutron matter provides a
perfect candidate for  dark matter.

As we have already described, the interaction is the 
Critical Pomeron. In the color superconducting theory (and before interactions
in QCD$_S$) the pomeron 
is a regge pole produced by reggeized gluon exchange
in the anomalous wee gluon condensate. This provides the basic interaction
which builds up the universal wee partons
needed for the parton model. Compared to conventional QCD, the QCD$_S$ states are 
fewer and the interaction is simpler - in agreement with experiment!
\begin{center}
{\l \it There is no BFKL pomeron, and no odderon.}
\end{center}
As part of our solution of the wee parton problem,
the non-perturbative physics of confinement and chiral 
symmetry breaking has a very simple 
``infinite momentum'' diagrammatic representation - supplemented
by an RFT critical phenomenon. 
Correspondingly, the transition from short-distance perturbation theory to
the regge region is also simple. In particular, the regge behavior of both the pomeron
and hadrons, which is very well established experimentally, is a direct
consequence of the regge behavior of gluons and quarks, which is similarly
well established theoretically.

We can not apply QCD$_S$, in isolation, to hadronic physics because
adding triplet quark masses 
would destroy the dynamics. Also, there will be a large
number of massless triplet quark mesons, which may well threaten the 
existence of an S-Matrix. 
The only way, that we know of, to add (effective) quark masses without destroying
the dynamics is to embed QCD$_S$ in QUD. Of course, as we have already described,
this embedding is also required by the addition of the electroweak interaction.

\mainhead{8. QUD } 

The construction of the bound-states and amplitudes of QUD is considerably
more complicated than the QCD$_S$ construction. The most essential new element of the
construction is the presence of
{\it fermion loop reggeon interactions} (without an anomaly)
that are {\it parity violating}. With a $k_{\perp}$-~cut-off, these interactions 
{\it exponentiate all anomaly divergences involving left-handed reggeon couplings}.
The conjugacy properties of QUD then imply that an anomaly divergence can occur only in 
an SU(3) subgroup and also that only the corresponding chirality transitions can be present.
Equivalently, the Dirac sea fluctuations producing wee gluon emissions are 
necessarily confined to a subgroup with charge conjugate 
fermion couplings. The result is the
{\it emergence of an SU(3) color singlet strong interaction}. The obvious question is,
of course, how is SU(5) invariance maintained? As we will see, the essential element
is the presence of the real SU(3) representation octet quarks
(which in going from QCD$_S$ to QUD seem, at first sight, to be unwanted!)

We follow the construction process described in Section 6.
Using the notation $SU(3)_C\otimes SU_L(2)\otimes U(1)$ to denote the decomposition 
of subsection {\bf 5.2}, we identify the various SU(5) subgroups as in Fig.~6. 
SU(3)$_C$ is, therefore, chosen to be a vector interaction. 
We restore the SU(2)$_C$ symmetry first, followed by the SU(4) symmetry.
Restoring the SU(3)$_C$ symmetry then coincides with the 
final restoration of the full SU(5) symmetry.

Because SU(2)$_C$ is a vector symmetry, it's restoration produces an
anomaly divergence which plays the central role
of selecting (what will eventually become) the physical states. As in QCD$_S$, 
these states are identified as anomaly pole Goldstone bosons ($~\pi_C$'s~)
with respect to some chiral symmetry that is present at this stage. 
These will be $~qq$, $\bar{q}\bar{q}$, and $q\bar{q}$ pairs 
in an SU(2)$_C$ condensate, 
where the $q$'s are {\bf 3's, 6's,} \& {\bf 8's} under SU(3)$_C$. Very importantly,
{\bf 8's} are real with respect to 
SU(3)$_C$, but contain complex doublets with respect to 
SU(2)$_C$ that will have the chiral symmetry required to produce anomaly pole
Goldstone bosons.
\begin{center} 
\epsfxsize=2.5in
\epsfbox{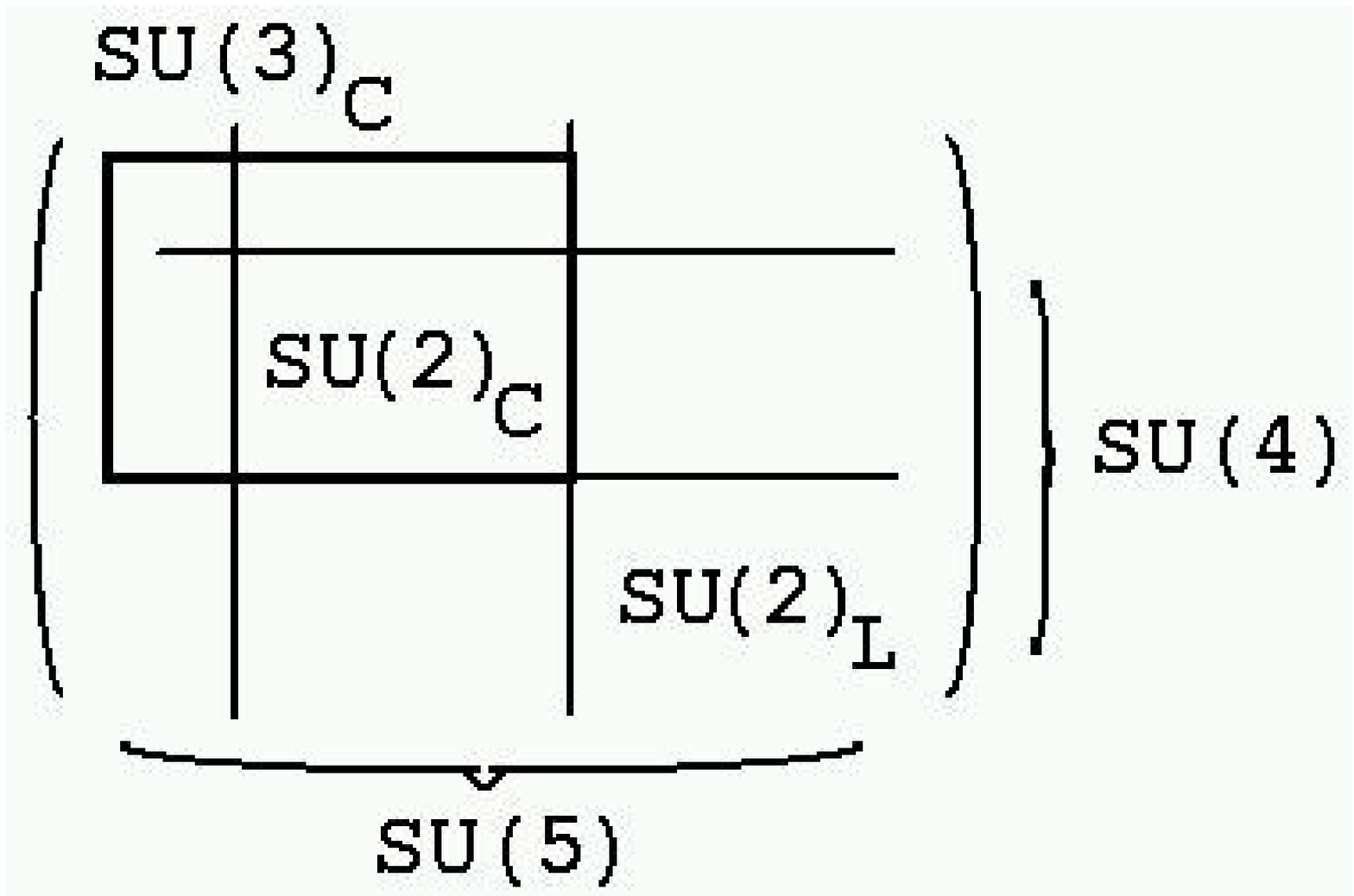}

Figure 6. SU(5) subgroups
\end{center}

An obvious interaction that will be SU(3)$_C$ invariant (after this symmetry is restored)
is an SU(2)$_L \otimes$U(1) singlet vector boson
in the SU(2)$_C$ condensate. This will become the pomeron. Interactions that
will similarly become SU(3)$_C$ invariant are the SU(2)$_L \otimes$U(1) electroweak
bosons - also in the condensate. The SU(2)$_C$ condensate additionally produces 
wee gluon anomaly interactions, of the kind illustrated as the last example
in Fig.~2, that 
\begin{center}
{\l \it give a mass to the left-handed electroweak bosons, 
\newline i.e. the $W^{\pm}$ and $Z^0$. $~~~~~~~~~~~$ } 
\end{center}
This process is equivalent
to a mixing with the $\pi_C$ states that, after the restoration of SU(3)$_C$, 
will be dominantly sextet pions. In the process, the $W^{\pm}$ and $Z^0$
aquire an anomaly-based flavor symmetry that protects them from an exponentiation that
would otherwise destroy the condensate that accompanies their exchange.

Restoring SU(4) symmetry 
involves only left-handed and abelian vector
bosons and so all new divergences exponentiate, leaving only   
states and interactions that are SU(4) invariant.
The  SU(2)$_C$ condensate can now be regarded as summed (or averaged) over all
SU(2) subgroups within SU(4), with SU(2)$_L$ defined always as the 
corresponding orthogonal SU(2) group. ``Leptons'' are present 
as reggeon bound states of ``elementary leptons'' and ``octet pions''
($\pi_C$'s composed of {\bf 8's})). ``Hadrons'' containing a triplet 
``pion'', or a triplet ``nucleon'', combined with octet quark pions
will also be present, as will the analagous sextet states.
The  SU(2)$_L \otimes$U(1) quantum numbers of octet $\pi$'s are such that 
the elementary lepton components must have (modulo gauge boson contributions) 
the generation structure of the Standard Model.
At this stage, both leptons and hadrons are formed via the same infra-red anomaly dynamics
and have essentially the same properties.

The restoration of SU(5) symmetry is an elaborate phenomenon
that, as yet, I only partially understand. 
In outline, the following is what I think happens. Firstly,
the pomeron becomes critical - as an SU(3) subgroup interaction
that is summed over subgroups. (Again, it is important that the color 
symmetries discussed 
here are global symmetries, and not gauge symmetries.) As in QCD$_S$, the 
lowest-order pomeron
is a vector reggeon exchange accompanied by a color-compensating (reggeon
condensate) shift in the Dirac sea 
within an SU(2) subgroup. To avoid divergence exponentiation,
the interactions that build up the Critical Pomeron are
necessarily confined to the same SU(3) subgroup.

As I currently understand it, the anomaly role of the octet quarks is at infinite momentum. 
With SU(3)$_C$ restored, the octet $\pi$'s 
no longer produce infra-red anomaly pole Goldstone bosons. Instead, as a consequence
of our cut-off manipulation, they 
contribute to bound-states as (what would normally be unphysical) zero mass anomaly
poles generated by on-shell quark/antiquark pairs which have infinite,
but opposite sign, energies. The cancelation of infra-red Goldstone boson 
effects has the consequence that the
chirality transition is transferred to infinite momentum. As a result, the creation of
SU(5) invariant physical states involves shifts of the 
Dirac sea at both infinite momentum (for SU(3) octets) and zero momentum 
(for SU(3) triplets and sextets). Therefore, as 
will be elaborated on in much more detail in \cite{amtm}, the
interplay between ultra-violet
and infra-red contributions to the triangle anomaly plays a crucial role in
the emergence of SU(5) invariance.

SU(3)$_C$ reality of the octet representation also 
implies that the octet $\pi$'s have no infra-red (anomaly) coupling to the 
pomeron and so leptons\footnote{A detailed description of lepton states is given
in my Fermilab talk listed in \cite{mct}.}
 have no strong interaction and no infra-red SU(3)$_C$ 
mass generation. With the octet $\pi$'s at large $k_{\perp}$,
the SU(2)$_L$ symmetry will appear, via the low $k_{\perp}$ components
of states, as the SU(2) flavor symmetry of the sextet sector. 
The quantum numbers of the octet $\pi$'s
should then produce the singlet/doublet structure of the Standard model.
If the bound-state lepton contribution
to the infra-red SU(2)$_L\otimes$U(1) anomaly has to be equal to the perturbative
contribution, the existence of three generations of leptons would be implied.
Anomaly cancelation would then
require three generations of quark hadrons that similarly
contain the octet pions.

After SU(3)$_C$ restoration the SU(2)$_C$ vacuum condensate contribution to
$\gamma,W^{\pm}$ and $Z^0$ exchange becomes an even signature 
``condensate'' effect that couples only to the large momentum
octet component of the states. (The infra-red
coupling to hadrons will be part of the interaction in which the
pomeron is exchanged along with the $\gamma,W^{\pm}$ and $Z^0$.)
There will also be a condensate coupling
to the octet quark component of the states
that will be part of the pomeron. Hence, the SU(5) invariant 
electroweak and strong interactions have very similar origins.
\begin{itemize}{\it 
\item{The electroweak interactions are perturbative reggeon exchanges
together with an orthogonal SU(3) singlet anomalous wee gluon exchange that 
couples via an octet Dirac sea shift at infinite momentum. Apart from
the mass generation, therefore, these interactions will effectively be perturbative
at low-energy.}
\item{The elementary pomeron (underlying the Critical Pomeron phenomenon)
is perturbative reggeon exchange together with
anomalous wee gluon exchange, both within an SU(3) subgroup, 
that couples via a triplet or sextet Dirac sea shift at
zero momentum and an octet shift
at infinite momentum.}}
\end{itemize}

The mass spectrum will be 
generated by a combination of
perturbative reggeization effects, 
color factors which (via the high mass sector, in particular) will emphasize the 
SU(3) strong interaction and
anomaly interactions  (analagous to that 
generating the $W$ and $Z$ masses) which will mix 
all the reggeon states. A wide range of scales will clearly emerge.
Without a better understanding of the anomaly interactions
and the related wee gluon distributions it is unclear, however, how many parameters 
may be involved.  
Although CP violation could be a consequence of
the anomaly dominance of interactions, with our current very limited
understanding of QUD it is certainly not obvious
that it is necessary. Nevertheless, given the absence of any conflicting symmetry,
there appears to be no reaon why the physical mass spectrum could not emerge.

\mainhead{9. A RADICAL PARADIGM CHANGE}

That the anomaly-dominated bound-state S-Matrix
of a very small coupling ($\alpha_u$ 
\raisebox{0.5mm}{${\scriptstyle <<}$} 1/50), massless, fixed-point, field theory 
is the origin of the Standard Model
is a radical proposal that could  
have many desirable consequences, if it is correct. We anticipate that the full field 
theory exists only perturbatively (at short distance) and that the S-Matrix  
scattering amplitudes are the only well-defined non-perturbative elements of the theory
(apart, perhaps, from the induced gravity referred to below). As we discuss next, it is 
essential for the physical applicability of our proposal that this be the case. 
The theoretical focus on the 
construction of a unitary particle S-Matrix within a perturbative field theory, 
rather than on the construction of a full
non-perturbative quantum field theory, could also be the change 
in paradigm that has been said
to be needed to end the ``crisis in fundamental physics''.
A priori, since massless fields and fermion anomalies are a crucial feature of the
dynamics of QUD, it would not be surprising if off-shell fields can not be constructed
for the bound-states we have described. 
The reggeon vertex anomalies that are at the core of our
amplitude construction are an S-Matrix phenomenon and, indeed, the general multi-regge
formalism has no off-shell parallel. Without this formalism, of course, 
there would be no possibility to discover the existence of a bound-state S-Matrix

In the current theory paradigm where, as described in Section 3, it is assumed
that (off-shell) field theory amplitudes are the essential elements of a gauge theory, 
both QCD$_S$ and QUD lie in what is referred to as the ``conformal window''. 
Within this window, the existence of an infra-red fixed-point
is anticipated\cite{asv}
to produce Green's functions with infra-red conformal properties that
would be inconsistent with a massive 
particle spectrum. It is presumably crucial, therefore, that QUD violate
the assumed paradigm by having no non-perturbative off-shell amplitudes. As a bonus(?),
the enormous theoretical 
challenge\cite{jw} of constructing a four-dimensional
non-perturbative off-shell gauge field theory would, of course, be avoided. Since 
physical masses and symmetries are bound-state S-Matrix properties, there is also
no need for a Higgs' field to generate masses or to break symmetries\footnote{As we 
have described, it is the chirality transitions in the anomaly vertices that, 
in effect, break symmetries in the S-Matrix.}. 
Therefore, the particular difficulty of constructing an asymptotically-free theory 
containing scalar fields is similarly avoided. One of the
lessons learnt from the focus on the S-Matrix during the bootstrap period 
was that (in the absence of quantized gravity, at least) 
the full physics content of a particle theory is contained within the S-Matrix\cite{hs}. 
Therefore, asymptotically-free perturbation theory 
together with a unitary bound-state S-Matrix 
within which an infinite momentum parton model is valid, could be
sufficient for all physical applications of QUD. Non-perturbative
off-shell amplitudes are a luxury that, from our perspective,
are not obviously necessary for any physical reason. 
Indeed, their absence may be a major factor
in unraveling many of the mysteries of the Standard Model.

\mainhead{10. CONSEQUENCES}

Obviously, an enormous part of the picture I have outlined 
remains to be established and it would be truly 
incredible if the Standard Model has the 
underlying simplicity that I am proposing. However, all the
necessary ingredients do seem to be present in QUD and, in strong contrast
with the epitome question of Section 2, it is
a unique theory which can not be added to or modified
in any way. If it is wrong, there is no 
fallback position and the conclusion has to be that the 
pursuit of a unitary S-Matrix, along the path I have 
followed, has simply been misguided. 

Although it will surely take a long time
to develop an in-depth understanding of all aspects 
of QUD, if my argumentation (or something not too different)
can be followed through it is a self-contained theory - with  
only Standard Model interactions. Assuming this is the case, there would be many 
benefits and many puzzles would be resolved.
\begin{enumerate} {\openup-0.5\jot \it
\item{The physics of the strong interaction (including confinement)
and the electroweak interaction is,
essentially, the same. Infinite momentum
non-perturbative physics 
is very close to perturbation theory and so
has a simple diagrammatic representation - with regge behavior 
emerging straightforwardly.}
\item{Parity conservation by the strong interaction and
parity violation by the weak interaction are naturally explained by the structure
of the anomaly interactions that result from the conjugacy properties of
the elementary fermion spectrum. (A mass is generated by anomaly interactions only for
left-handed gauge bosons.)}
\item{The only new physics is the strong interaction color sextet 
sector. This sector simultaneously explains the two major mysteries of the Standard Model, 
namely the origin of electroweak symmetry breaking and the nature of dark matter. 
The sextet meson and baryon sectors separately provide what 
is, surely, a remarkably economic resolution of these two mysteries. 
There is
already suggestive experimental evidence\cite{mct,arw05}
that this is, indeed, the right ``new physics''.}
\item{The underlying gauge symmetry implies that there will be unification of the
couplings. The presence of the high-mass strong interaction sector allows
this to be achieved without supersymmetry!} 
\item{The mass spectrum is not simply determined by elementary masses and the
scale evolution of the couplings. Wee parton
anomaly interactions mix the reggeon states and, presumably, 
introduce parameters. The color factors involved will produce a wide range of 
scales and masses that could very well produce the Standard Model spectrum, since
there is no conflicting symmetry.}
\item{The smallness of the lightest particle (neutrino) masses should be a direct
reflection of the small coupling in the underlying field theory.}
\item{There is no proton decay, but 
the experimentally
attractive SU(5) Weinberg angle should hold!}
\item{Because QUD is an asymptotically free, massless, fixed-point theory, 
it has no vacuum energy and (in the absence of the quantization of gravity)
would\cite{bh} (perturbatively) induce Einstein gravity with 
zero cosmological constant.}}
\end{enumerate}
It should be mentioned that reggeon unitarity could be an insuperable problem
for the quantization of gravity. If  
the graviton appears as a reggeized particle in an S-Matrix, reggeon unitarity implies that
the exchange of arbitrary numbers of reggeized gravitons would produce non-polynomially
bounded S-Matrix amplitudes (a non-local theory?). 
This is, perhaps, a major argument against the quantization of gravity.

An underlying field theory for the Standard Model
in which no fields have mass is surely attractive, both
theoretically and aesthetically. With regard to the possible origin of the QUD
fermion representation it could be significant that it is contained in a single 
anomaly-free SO(10) representation - the {\bf 144}. 

\mainhead{11. WHAT SHOULD BE SEEN AT THE LHC?}

Since our proposal is so radical, it is fortunate that there are major experimental
predictions for the LHC, for some of 
which it will be hard not to acknowledge their significance - if they are seen.  
Large cross-section effects are expected that should 
make the emergence of the sextet sector obvious. If these effects are not seen then
either my understanding of QUD is wrong or the theory is irrelevant!
The following is a very brief review of what is discussed in 
much more detail in \cite{arw05} and is also discussed in \cite{mct}. 

Most immediately, multiple vector boson and jet x-sections will be
much, much, larger than expected, with $<p_{\perp}>$  
approaching the electroweak scale. For these phenomena
there will, however, be competing explanations, e.g. black holes, sphalerons, etc.

A priori, $N_6\bar{N}_6$ pair production (dark matter) should be seen - with 
$m_{N_6} \sim 500 ~GeV$ ? But, missing energies of several hundred GeV will be common
and the low energy $N_6$ hadronic 
cross-section (in a calorimeter) is probably small.
$P_6\bar{P}_6$  pair production should be seen
(if the $P_6$ is not too unstable). Again, though, will
a massive charged particle with a large production x-section be seen as
direct evidence for the sextet sector? 

The double pomeron cross-section could provide 
the most definitive early evidence for the sextet sector.
With the pomerons detected via Roman pots, the environment is clean.
$W$ and $Z$ pairs will be produced in the double pomeron 
cross-section via sextet pion anomaly poles and so, 
as pion pairs dominate the double pomeron
cross-section at low mass, so $W$ and $Z$ pair production will dominate the
cross-section at the electroweak mass scale.

When $|k_{\perp}|$ is electroweak scale,
the double-pomeron $ W$ and $Z$ pair amplitude for producing jets is comparable
with a standard jet amplitude that has, apart from anomaly
loops that are presumably O(1), the same propagators and couplings. This suggests
that the jet cross-section from double-pomeron $W$ and $Z$ pairs, produced as 
illustrated in Fig.~5,
\begin{center} 
\epsfxsize=4in
\epsfbox{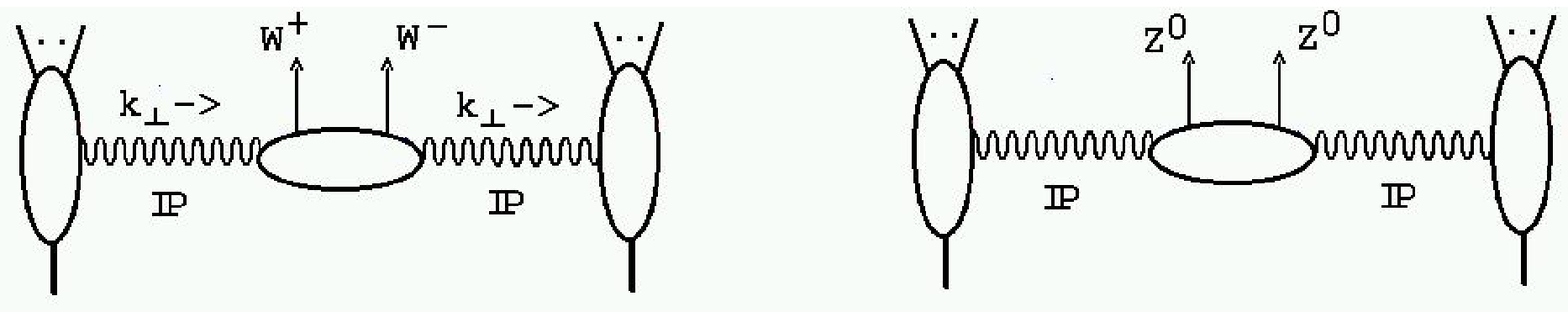}

Figure 7. Double Pomeron Amplitudes 
\end{center}
 will be comparable with the non-diffractive jet
cross-sections predicted by standard QCD.
While the $~\pom~ W^+W^-~\pom~$ and $~\pom~ Z^0Z^0~\pom~$
vertices appearing in Fig.~5 should vary only slowly with $k_{\perp}$, the $pp~\pom~$ 
vertices have strong $ k_{\perp}$-dependence. This implies there should be
an extremely large x-section at small $t$. 

In the initial low luminosity running, an ``extremely large x-section''
could be detected by TOTEM/CMS.
There could be spectacular events in which protons are tagged and only (a multitude of)
large $E_T$ charged leptons are seen in the central detector.
FP420 will take over during the high luminosity running 
and should surely see an enhanced cross-section, even if it is too small 
to have been seen by CMS/TOTEM. Most likely,
the $W$ and $Z$ pair cross-section will overwhelm all other physics.

A large double-pomeron cross-section for $W$ and $Z$ pairs
implies that the longitudinal components of the $W$ and the $Z$  
have direct strong interactions. The only known possibility for this
is the existence of the sextet sector and, as we have discussed, to give a well-defined 
theory this sector has to be embedded in QUD !

After $\pom$, $W/Z$, and jet physics has established that 
sextet quark physics is definitively discovered, 
the search for ``dark matter'' will become all important.
The cross-section for double-pomeron production
of stable $N_6\bar{N}_6$ pairs (with a pair mass
$~ \centerunder{\raisebox{0.5mm}{${\scriptstyle >}$}}{${\scriptstyle \sim}$}
~1~TeV$) 
could be large enough that it will be definitively seen by the forward pot
experiments. It will be a spectacular process to look for - via the following.
\begin{enumerate}
\item {The tagged protons determine
a very massive state is produced.}
\item {No charged particles are seen
in any of the detectors.}
\item{Having low energy, the $N_6$ hadronic 
cross-section will, probably, be small
but some hadronic activity may be seen in the central calorimeter.}
\item {Charged lepton comparison would 
allow a separation with respect to the multiple $Z^0$ production of neutrinos.}
\end{enumerate}
If the $P_6$ is relatively stable, 
and not too different in mass from the $N_6$ it would, of course, be much simpler
to first detect $P_6\bar{P}_6$ pairs.

In \cite{arw05} we gave a lengthy discussion of the possibility that deep-inelastic
diffractive $Z^0$ production (via the longitudinal sextet pion
component of the $Z^0$) could be responsible for events with very large $x$
and $Q^2$ at HERA. DIS diffractive $Z^0$ production could also be\cite{mga} an important
discovery process for the sextet sector at the LHC and, perhaps, even at the Tevatron.

\end{document}